\documentclass{JINST}
\usepackage{graphicx}

\newcommand{\lr}[1]{\left(#1 \right)}

\title{Spurious shear induced by the tree rings of the LSST CCDs}

\author{Yuki Okura$^a$$^b$,
Andr\'es A. Plazas$^c$$^d$,
Morgan May$^c$,
 Toru Tamagawa$^a$\\
\llap{$^a$}RIKEN,\\
  2-1 Hirosawa, Wako, Saitama 351-0198, Japan\\
\llap{$^b$}RIKEN-BNL Research Center,\\
 Department of Physics, Brookhaven National Laboratory, Bldg. 510, Upton, NY, 11792, USA\\
\llap{$^c$}Brookhaven National Laboratory,\\
 Department of Physics, Brookhaven National Laboratory, Bldg. 510, Upton, NY, 11792, USA\\
\llap{$^d$}Jet Propulsion Laboratory,\\
 California Institute of Technology, 4800 Oak Grove Drive, Pasadena, CA, 91109, USA\\
E-mail: \email{yuki.okura@riken.jp}}

\abstract{We present an analysis of the impact of the tree rings seen in the candidate sensors of the Large Synoptic Survey Telescope (LSST) on galaxy-shape measurements. The tree rings are a consequence of transverse electric fields caused by circularly symmetric impurity gradients in the silicon of the sensors. They effectively modify the pixel area and shift the photogenerated charge around, displacing the observed photon positions. The displacement distribution generates distortions that cause spurious shears correlated with the tree-rings patterns, potentially biasing cosmic shear measurements. In this paper we quantify the amplitude of the spurious shear caused by the tree rings on the LSST candidate sensors, and calculate its 2-point correlation function. We find that 2-point correlation function of the spurious shear on an area equivalent to the LSST field of view is order of about $10^{-13}$, providing a negligible contribution to the 2-point correlation of the cosmic shear signal. Additional work is underway, and the final results and analysis will be published elsewhere (Okura et al. (2015), in prep.)
}


\begin{document}

\section{Introduction}\label{sec:Intro}
One of the most important studies in current cosmology is the constraint of the parameters of the standard cosmological model, including the equation of state of dark energy. Weak lensing is recognized as one of the most powerful methods to constrain the current cosmological model through the characterization of the mass distribution and evolution of the large scale structure (LSS) of the universe(\cite{Kaiser2000}, \cite{Wittman2000}, \cite{Miller2007}, \cite{Schneider2006}, \cite{Munshi2008}, \cite{Kilbinger2014}).  

As a consequence, several galaxy surveys that use the weak lensing of the LSS as one of their main techniques have started to operate (e.g. Dark Energy Survey\footnote{http://www.darkenergysurvey.org/}, HSC\footnote{http://www.naoj.org/Projects/HSC/HSCProject.html}, PanStarrs\footnote{pan-starrs.ifa.hawaii.edu/}, KIDS\footnote{http://kids.strw.leidenuniv.nl/}, etc.) or are being planned (e.g, LSST\footnote{http://www.lsst.org}, EUCLID\footnote{http://sci.esa.int/euclid}, WFIRST-AFTA\footnote{http://wfirst.gsfc.nasa.gov/}). The cosmic shear analysis relies on the accurate measurement of shapes of millions or up to billions of background galaxies, to statistically estimate the small distortion (about $0.1\%$) caused by all the large scale structure in the universe. Therefore, it is crucial that all systematic errors comparable or larger than this threshold be corrected in order to accurately infer cosmological parameters. 

One of the instrumental features that has been identified in candidate CCDs for LSST and in detectors of other cameras (such as the Dark Energy Camera) is called tree rings (\cite{plazas2014}, \cite{Stubbs2014}, \cite{Lupton2014}, \cite{Jarvis2014}). Under uniform illumination, these appear as concentric circles whose center coincide with the center of the silicon wafer from which the detectors were cut. The origin of this effect lies in impurities in the silicon that generate spurious fields parallel to the surface of the CCD, transverse to the main electric field. The effective area of the pixels is modified and the photogenerated charges shifted around, generating biases in photometric and astrometric measurements, if not corrected. In particular, the spatial variation of the astrometric displacements distorts the shapes of objects, producing a coherent spurious signal that mimics the coherent distortion produced by weak lensing. Therefore, it is important to characterize the amplitude of the shear induced by the the tree rings, to determine their impact on cosmological parameter estimations from weak lensing. 

This paper is organized as follows: in Section 2, we explain how we can estimate spurious shear from a concentric displacement such as the one produced by the tree ring effect. In Section 3, we measure the spurious shear and its 2-point correlation by using  two types of candidates CCDs for LSST CCDs. We summarize our findings and discuss further work in Section 4.

\section{Distortion by Concentric Displacements}\label{sec:DCD}
\cite{plazas2014} (hereafter PBS14) used images with uniform illumination (dome flat-field images, or flat fields) to characterize the tree rings of the 62 science CCDs of the Dark Energy Camera (DECam) \cite{diehl2012} focal plane. After identifying the center of the tree rings in each CCD, radial profiles of their amplitudes as a function of distance from that center were created, assuming azimuthal symmetry. These profiles were directly used to quantify the photometric and astrometric displacements caused by the tree rings, and actual data from the science verification period of DES were used to constraint the astrometric displacements by the tree rings. The astrometric and photometric templates were incorporated into the photometric and astrometric (WCS) solutions to achieve residuals below the scientific requirements for DES. In this section we show how these kinds of displacements lead to distortions in the shapes of measured objects. 

\subsection{Spurious Shear estimated from Concentric Displacements}\label{sec:SSCD}
\begin{figure}[tbp] 
\centering
\includegraphics[width=.5\textwidth]{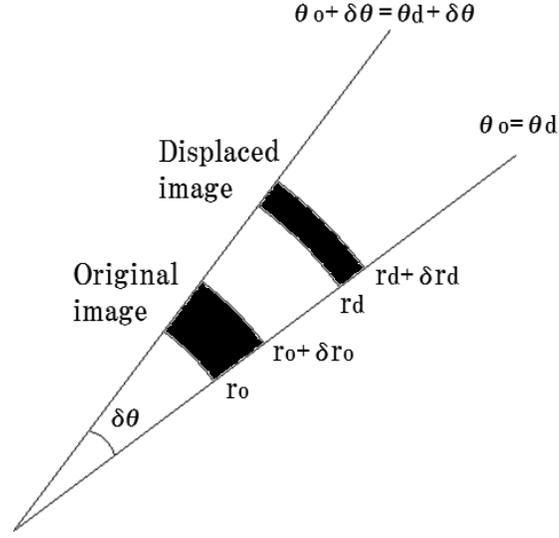}
\caption{Concentric displacement of an infinitesimal area by a distortion $d(r)$.}
\label{fig:CD}
\end{figure}

Let us consider a displaced infinitesimal image between radius $r_d$ and $r_d+\delta r_d$ and $\theta_d$ and $\theta_d+\delta \theta_d$ in polar coordinates, where the origin of the coordinate system is set to the center of the concentric displacement. The image is shifted from the original position $r_o$ and $r_o+\delta r_o$ and $\theta_o$ and $\theta_o+\delta \theta_o$ by the tree ring displacement, which is written as a function $d(r_d)$. Figure \ref{fig:CD} shows the relation between the images. Then the edges of the image in the $r$ direction relate to each other as follows:
\begin{eqnarray}
r_o&=&r_d-d(r_d)\\
r_o+\delta r_o&=&r_d+\delta r_d-d(r_d+\delta r_d).
\end{eqnarray}
The distortion can be obtained by taking the ratio of the length of the images in both the radial and tangential direction. Thus, the radial distortion is given by
\begin{eqnarray}
\label{eq:cdist_r}
\frac{r_o+\delta r_o - r_o}{r_d+\delta r_d - r_d}=1-\frac{d(r_d+\delta r_d) - d(r_d)}{\delta r_d}=1-\frac{\partial d(r)}{\partial r}  \bigg |_{r=r_d},
\end{eqnarray}
and the tangential distortion is given by
\begin{eqnarray}
\label{eq:cdist_t}
\frac{r_o\delta \theta}{r_d\delta \theta}=1-\frac{d(r_d)}{r_d}.
\end{eqnarray}
For convenience, and without loss of generality, let us consider a situation where $\theta_o=0$. Then we can arrange Eqs. \ref{eq:cdist_r} and \ref{eq:cdist_t} in matrix form, obtaining
\begin{equation}
\left(
\begin{array}{c}
\delta r_o\\
r_o\delta\theta
\end{array}
\right)=\left(
\begin{array}{cc}
1-\frac{\partial d(r)}{\partial r}|_{r=r_d}&0\\
0&1-\frac{d(r_d)}{r_d}
\end{array}
\right)\left(
\begin{array}{c}
\delta r_d\\
r_d\delta\theta
\end{array}
\right) \\ 
\equiv\left(
\begin{array}{cc}
1-\kappa^{TR}-\gamma_{rad}^{TR}&0\\
0&1-\kappa^{TR}+\gamma_{rad}^{TR}
\end{array}
\right)\left(
\begin{array}{c}
\delta r_d\\
r_d\delta\theta
\end{array}
\right)
\end{equation}
where $\kappa^{TR}$ and $\gamma^{TR}$ are the convergence and shear due to the tree-rings displacement, defined as
\begin{eqnarray}
\label{eq:conv_d}
\kappa^{TR}(r)&\equiv&\frac12\lr{\frac{\partial d(r)}{\partial r}+\frac{d(r)}{r}}\\
\label{eq:shear_d}
\gamma^{TR}_{rad}(r)&\equiv&\frac12\lr{\frac{\partial d(r)}{\partial r}-\frac{d(r)}{r}}
\end{eqnarray}
Thus, the two components of the shear, $\gamma_1$ and $\gamma_2$, at $(r,\theta)$ can be written as
\begin{eqnarray}
\label{eq:shear_12}
\gamma^{TR}_{1}(r,\theta)+i\gamma^{TR}_{2}(r,\theta)&=&\gamma^{TR}_{rad}(r)e^{2i\theta}
\end{eqnarray}

\subsection{Spurious Shear estimated from Flux Modulation}\label{sec:SSFM}
In practice, it is posible to directly measure the astrometric displacement $d(r)$ caused by the tree rings by using sets of several dithered exposures of star fields in different photometric bands and constructing ``star flat" images (\cite{Manfred1995}, \cite{Tucker2007}). However, this is not an easy task given that the displacement is small (of the order of subpixels). PBS14 demonstrated that, to first order, there is an relationship between the tree-rings flux modulation $f(r)$ as measured by the flat fields (which have higher S/N compared to the one provided by the finite number of stars in the star flats) and the astrometric displacement $d(r)$, given by 
\begin{eqnarray}
\label{eq:d_f}
d(r)=-\frac{1}{r}\int_0^r dr' r'f(r'),
\end{eqnarray}
The minus sign comes from the difference between the definitions of $d(r)$.
By using Eq. \ref{eq:d_f}, we can rewrite Eqs. \ref{eq:conv_d} and \ref{eq:shear_d} as
\begin{eqnarray}
\label{eq:conv_f}
\kappa^{TR}(r)&=&-\frac12f(r)\\
\label{eq:shear_f}
\gamma^{TR}_{rad}(r)&=&-\frac12\lr{f(r)-2\frac{d(r)}{r}}\approx-\frac12f(r),
\end{eqnarray}
where the last approximation is valid when the displacement is much smaller than the radius. The typical scales of the spatial displacements are of the order of subpixels, and the radii of the order of 100$\sim$1000 pixels, so the approximation is valid. 
This means that the convergence and shear due to the tree rings can be estimated as approximately half the value of the flux modulation.

\section{Spurious Shear and 2-point Correlation Function from tree rings on LSST CCDs}\label{sec:SS2CF}
In this section, we estimate the spurious shear caused by the tree-ring effect on two types of candidate CCDs for the LSST project by using Eq. \ref{eq:shear_f}. It is important to keep in mind that this relation is true for an infinitesimal region, and therefore more realistic effects such as brightness distribution, pixelization, and PSF convolution need a more careful treatment. In particular, the PSF correction in weak lensing shear analysis should not be neglected when estimating the spurious shear by the tree rings, because the tree-ring effect changes the ellipticity of the galaxies after PSF smearing. Therefore, in most cases, the PSF correction augments the spurious shear caused by the tree rings, e.g. in the situation of a circular PSF. However, in this paper we only estimate the impact on shear measurements due to the tree rings in a more simplified situation
\subsection{Flux modulation, displacement, and spurious shear by the tree rings}\label{sec:MDS}
We measured the tree-ring patterns of two different LSST candidate CCDs individually. Each one is fabricated by a different vendor, and we refer to them as ``type 1" and ``type 2" CCDs. We used flat-field images from a uniform light source of $750$ nm wavelength. These flat images have a 4k $\times$ 4k pixel size and a standard deviation about 0.4\%.

First we corrected the flat field images to eliminate other effects. We masked regions near the edge of the detectors and shadows due to dust particles in the optics. The light source in this test was not completely uniform, so we fit the slow variation with a 7th order polynomial.  Other sensor effects are apparent in the image after correction due to pixel size variations and laser annealing. Though pixel size variation also will result in its own pattern of spurious shear, we removed both these effects by scaling the image. Though not ideal, this treatment is adequate for the present purpose. In future analysis we will study these effects in more detail.

Next, we determine the center of the tree ring profiles in each CCD. We do this by selecting points on several rings and then fitting for their common center. The centers of the rings for type 1 and 2 CCD lie at [4575,-375] pix [-167,4172] pix, respectively, where the origin of the coordinate system is at the bottom left corner of the CCD (see Figure \ref {fig:SSTR_2dim_color_CCD_Type1} for one of the CCD types). The position is near the corner, but slightly offset from it. 

\begin{figure}[tbp] 
\centering
\includegraphics[width=\textwidth]{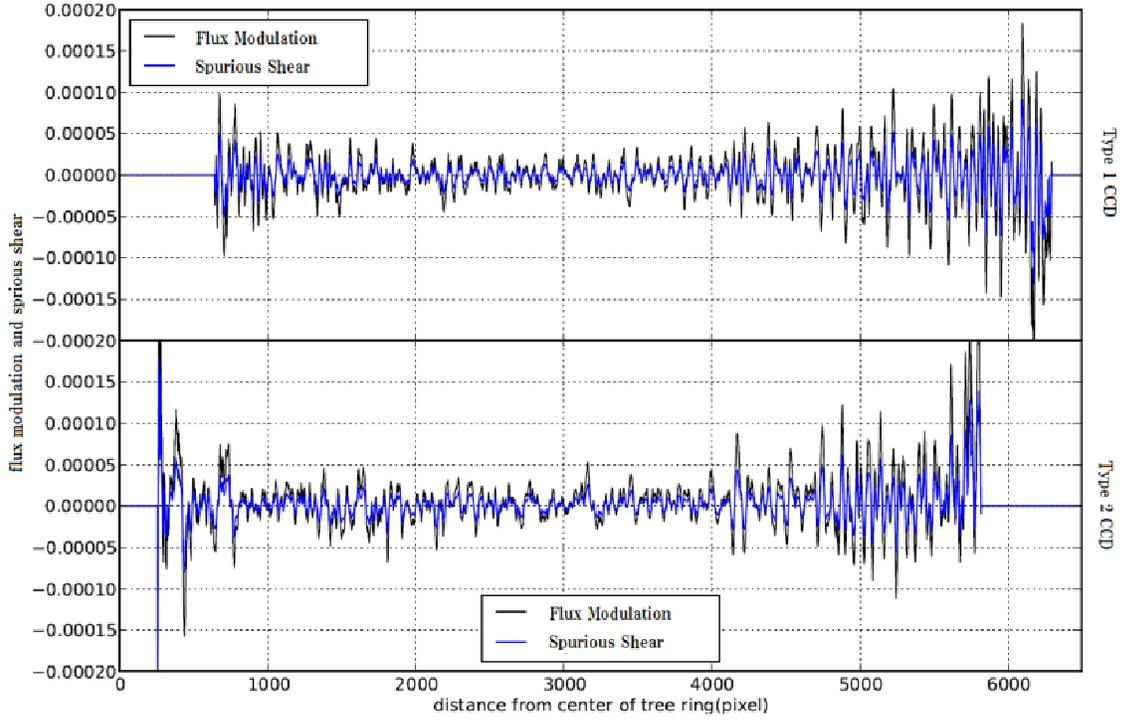}
\caption{1-dimensional profile of the flux modulation(black) and the spurious shear(blue) caused by the tree rings effect on type1(top) and type 2(bottom) LSST CCDs. The tangential shear is estimated from the flux modulation by using Eq. 2.11.
}
\label{fig:SSTR_1dim}
\end{figure}

The two panels in Fig. \ref{fig:SSTR_1dim} show the measured tree-ring profiles of type 1 and 2 CCDs, respectively. 
Before measuring the profiles, the images were smoothed by a Gaussian function with a 5-pixel kernel. Since the scale of width of the tree rings peaks is of about $10$ pixels, the smoothing should not reduce their amplitude by much. The typical scale of the flux modulations is approximately $0.01\%$, about 50 times smaller than that on the DECam CCDs (PBS14). The blue lines in Fig. \ref{fig:SSTR_1dim} represent the tangential spurious shear calculated from the flux modulations by using Eq. \ref{eq:shear_f} (the spurious shear is about half value of flux modulation). The typical amplitude scale of the spurious shear is of about $0.005$\%.

\begin{figure}[tbp] 
\centering
\includegraphics[width=\textwidth]{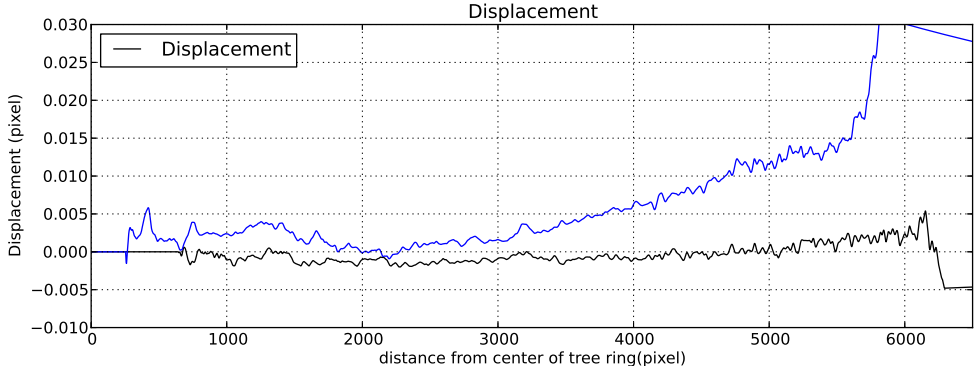}
\caption{1-dimensional profile of the displacement caused by the tree-ring effect on type1 (black) and 2 (blue) LSST CCD.
}
\label{fig:SDTR_1dim_Type12}
\end{figure}

Fig. \ref{fig:SDTR_1dim_Type12} shows the astrometric displacement by the tree rings on both types of CCDs. The displacement is calculated by using Eq. \ref{eq:d_f}, which is an integral equation over the flux modulation. Therefore, large fluctuation errors from noise remain at large distances. Such errors could be corrected by using measured astrometric displacement from real stars (\cite{plazas2014}). However, we can say that the typical amplitude is of about 0.001 pixels.

\subsection{2-point Correlation Function}\label{sec:2CFLSST}
\begin{figure}[tbp] 
\centering
\includegraphics[width=\textwidth]{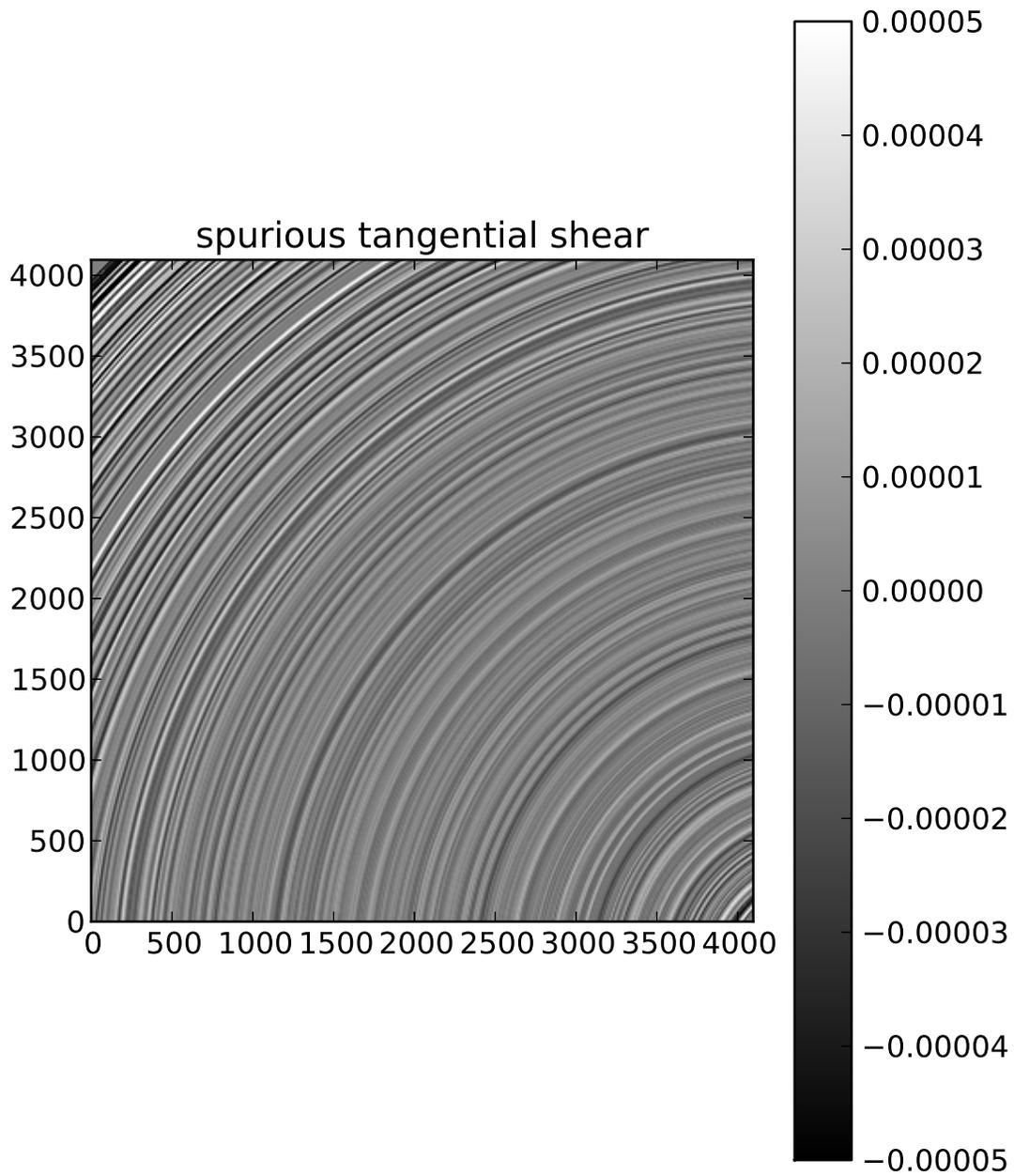}
\caption{Tangential spurious shear produced by the tree rings on type1 LSST CCD, calculated from the flux modulation measured from the flat fields}
\label{fig:SSTR_2dim_color_CCD_Type1}
\end{figure}

We now present the 2-point correlation of the spurious shear induced by the tree-ring effect on the LSST CCDs.  First, we calculated the spurious shear pattern on each individual CCD (Fig.\ref{fig:SSTR_2dim_color_CCD_Type1}). The objects(galaxies) we observe have finite size, so the spurious shear for them is not the same as the effect on pixel scales. However, since the variation scale of the tree rings is larger than the object scale, we assume that the spurious shear effect does not change over the object scale.Therefore, the spurious shear for objects is almost the same as the effect on pixel scales, so we have calculated the 2-point correlation of the spurious shear for the latter.

\begin{figure}[tbp] 
\centering
\includegraphics[width=\textwidth]{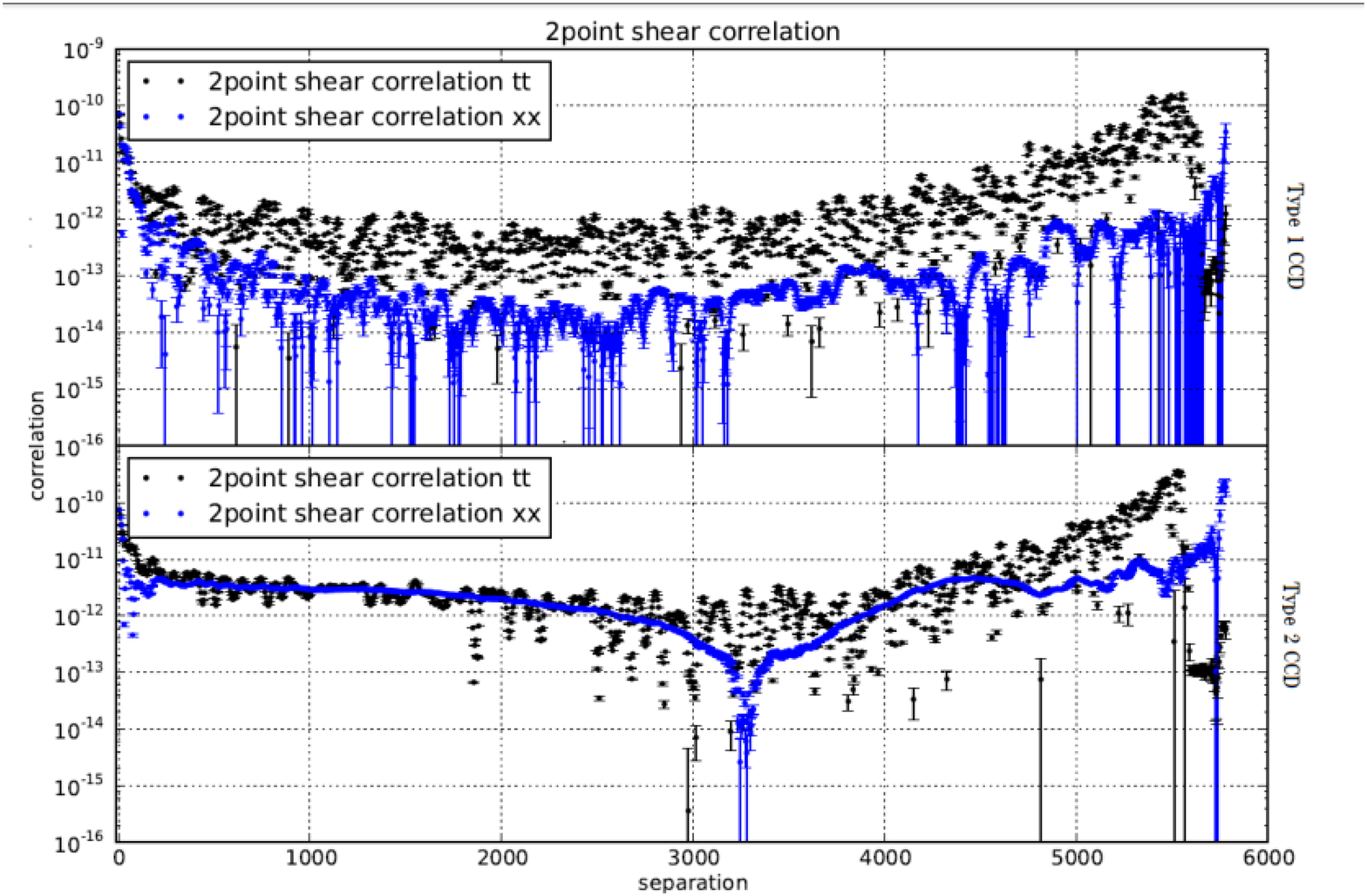}
\caption{Absolute values of the 2-point spurious shear correlations caused by the tree-ring effect on type 1(top) and type 2 (bottom) LSST CCD. Black points mean parallel correlation and blue points mean cross correlation with 5 pixels sampling. The values oscillate around zero.
}
\label{fig:SSTR_2cor_CCD_log}
\end{figure}

Fig. \ref{fig:SSTR_2cor_CCD_log} shows the 2-point correlation function of the spurious shear on both CCD types. The typical value is about $10^{-11}$, which is small, although at very short spatial scales we can see a slightly larger peak. This value is much smaller than the square of the typical scale of the spurious shear, $(5\times10^{-5})^2 \approx 10^{-9}$. The reason for the smallness is that the tree-ring oscillation gives images of objects an apparent spurious  tangential and radial ellipticity in an alternate fashion, and therefore when calculating the correlation function the values cancel each other out. 


\begin{figure}[tbp] 
\centering
\includegraphics[width=\textwidth]{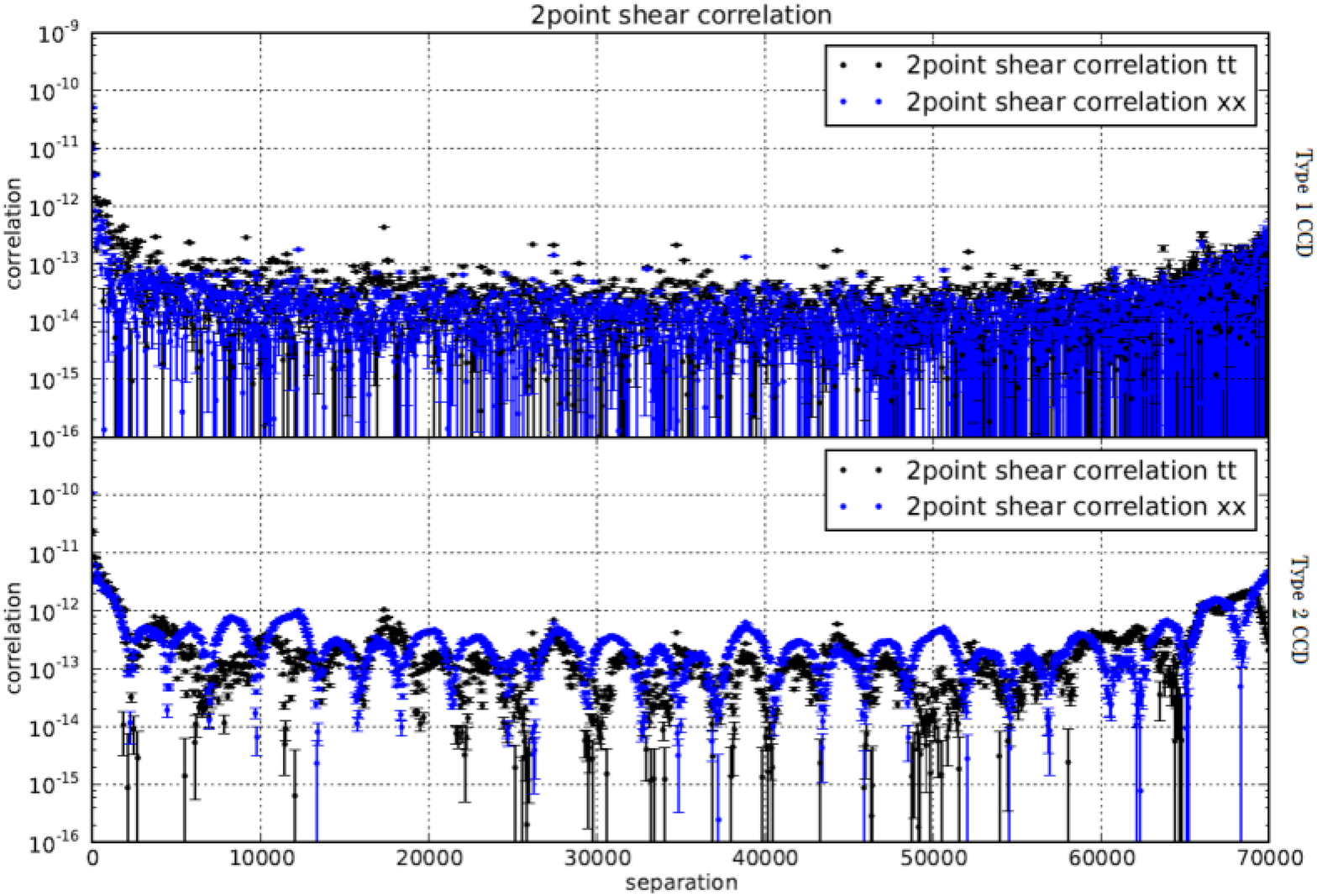}
\caption{Absolute values of 2-point spurious shear correlation caused by the tree ring effect on FOV of Type 1(top) and Type 2(bottom) LSST CCD. Black points mean parallel correlation and blue points mean cross correlation with 50pixels sampling. The values oscillate around zero.
}
\label{fig:SSTR_2cor_FOV_log}
\end{figure}

Fig. \ref{fig:SSTR_2cor_FOV_log} shows the 2-point correlation of the spurious shear on type 1 and type 2 over an area equivalent to the full field of view of the LSST camera with a $100$ pixel sampling scale. The typical scale for the amplitude decreases to $10^{-13}$. In this calculation, we have assumed that all CCDs in the raft have the same tree ring profile but with a different orientation, since we have only tested one CCD of each type and four squares of silicon are cut out of the circular cross section of silicon, in such a way that the center of the boule, and of the tree ring pattern is at a corner of each square. Corners which have the center of tree-ring pattern are not same in all CCDs.

\section{Summary}\label{sec:Sum}
We have presented a study of the spurious shear induced by the astrometric displacement caused by the tree rings in LSST candidate CCDs. \cite{plazas2014} (PBS14) studied the tree rings on the DECam CCDs, and presented a model that estimates the astrometric displacement from the flux modulations by the tree rings as measured in flat-field images. Inferring the astrometric displacements in this way is relatively easier than direct measurements of the astrometric residuals from real data. 

We developed new formulae that calculate the spurious shear from the astrometric displacements. By using the fact that the tree-ring effect induces a concentric displacement, the spurious shear can be estimated from the derivative of the displacement function, where the displacement function is calculated from the flux modulation. We then measured the tree ring pattern on two LSST candidate CCDs, and calculated the induced shear. The typical value of flux modulation due to tree rings is $0.01\%$, about 50 times smaller than that for the DECam CCDs. In addition, the typical amplitude of the 2-point correlation function in the field of view of the LSST is about $10^{-13}$, which is small enough to be neglected in calculations that constrain cosmological parameters. The correlation is much smaller than the square of the typical amplitude of the spurious shear due to cancelations of tangential and radial shears in the calculation.

In this study we have measured only the impact of tree rings on the 2-point correlation function, but  we plan to extend our analysis to include other sensor anomalies and the impacts of tree rings on cosmological parameters.

\acknowledgments
We thank G. Bernstein and P. O'Connor for useful comments and discussions. 
This work was supported in part by the U.S. Department of Energy under Contract No. DE-AC02-98CH10886 and Contract No. DE-SC0012704.
AAP is also supported by JPL, which is run under a contract for NASA by Caltech.

\end{document}